# Periodic steps in the resistance vs. temperature characteristics of doped graphite and graphene: evidence of superconductivity?

**Grover Larkins, Yuriy Vlasov and Kiar Holland**

Department of Electrical and Computer Engineering, College of Engineering, Florida International University, 10555 W. Flagler Street, Miami, Florida 33174

E-mail: vlassovy@fiu.edu

**Abstract**
We have observed periodically repeated steps in the resistance vs. temperature characteristics of doped Highly Oriented Pyrolytic Graphite and exfoliated doped multi-layer graphene. The observations consist of a series of regularly spaced steps in the resistance vs. temperature curves. The lowest step is observed at a temperature of from 50 to 60 K. Additional steps are observed at multiples of that basic temperature with the highest step temperature being at approximately 270 K. Quenching by a modest applied magnetic field has been observed. The sizes and widths of the observed steps appear to vary and may be related to some sort of aggregation reminiscent of flux vortex pinning. An additional argon ion implantation at reduced energy was done to see if additional defects in the material would yield additional structure. This was observed to yield much sharper resistance steps at temperatures in excess of 200 K in thin exfoliated peels from the implanted surface of the sample and would support the possibility of the steps being related to pancake vortex pinning in a layered superconductor. Unfortunately, as yet, there has been no direct measurement of either the superconducting energy gap nor has a definitive Meissner effect been observed.

**1. Introduction**
Carbon has large number of forms and structures. All of these exhibit different mechanical, chemical and electrical properties [1]. Graphene [2], a two-dimensional carbon structure, has been suggested for use as a test object to study possible electron pairing and superconductivity. Highly Oriented Pyrolytic Graphite (HOPG) with its' weak interlayer coupling [3,4] can be used in this study since it approximates graphene fairly well. Our earlier work [5] has given every indication that there is superconductivity in doped HOPG and, possibly, graphene as well. For this reason it was decided to extend the work to include graphene, both exfoliated from ion implanted HOPG and doped, *in situ*, grown graphene as a subject of study for superconductivity.

The observations reported here in this work do not fit any phenomenon known to us other than superconductivity. Although the evidence is not conclusive that superconductivity is the only explanation for the observed behaviour, given the lack of any other known model to fit the data within, we have chosen to follow the possibility that superconductivity is present. It is important to note that we are not ruling out some new physical manifestation of the extreme anisotropy and two dimensionality of the material but that we have no model other than superconductivity to base any behavioural predictions on.

## 2. Background

It has been hypothesized that the close coupling or strong scattering of electrons in graphene indicates the potential for superconductivity to exist at, or near, room temperature in doped graphene [5-10]. Kopelevich *et al.* [11-13] have reported some cases of possible superconductivity in Highly Oriented Pyrolytic Graphite but, given that their results were only on as-grown samples, the origins and likelihood of the observations being due to superconductivity were unclear. Our prior work [5] is the first that we are aware of where a systematic attempt has been made to substitutionally dope HOPG/graphene/graphite into a superconductive state. Additional, later, work by Scheike *et al.* [14,15] and Ballestar *et al.* [16] has also given hints of possible superconductivity in doped graphite. The work described herein represents our effort to attempt to confirm or disprove these results as originating from superconductivity.

Moving from bulk ion-implanted HOPG to multilayer graphene exfoliated from ion-implanted HOPG allowed our measurements to remove bulk effects from the physics occurring in the first 20 nm where the dopant and damage from the implantation were located.

A question that arose from our work [5] on doped HOPG was why the resistance remained non-zero at low temperatures. A review of the literature involving resistance measurements in layered superconductors leads to the immediate identification of flux-flow resistance as a potential reason for the non-zero resistance at low temperatures [17,18]. The literature indicates that flux pinning is a key in reducing flux-flow resistance in the layered high-$T_c$ superconductors [17-25] and would be an essential factor here as well should this be due to vortex flow. Another point that becomes clear upon examination is that, with an exfoliated film only a nanometer or two thick, the test current in a resistance vs. temperature measurement must be as small as possible or it could potentially influence the results.

An examination of the volume of prior work [19-26] that has been done to characterize and model the behaviour of magnetic vortices in layered superconductors leads to the following general conclusion: Pancake vortices would be the preferred vortex form if the material is superconducting. A more specific conclusion (derived from the work by Pe, Benkraouda and Clem [21]) is that for graphene stacks with interlayer distances of 0.2 nm and relatively long magnetic penetration depths (graphene/graphite are diamagnetic materials with significant magnetic anisotropy [27]) that the interlayer coupling, if the material was a superconductor with a magnetic self pinning attractive force between pancake vortices in layers *i* and *j* given as follows:

$$\mathbf{F}(\mathbf{r}_j, j, i) \approx -\hat{\rho}(\phi_j)\left(\frac{\varphi_0}{2\pi\Lambda}\right)^2 \frac{\sqrt{\rho_j^2 + |i-j|^2 s^2} - |i-j|s}{\rho_j} \quad (1)$$

$$\text{where } \Lambda = 2\lambda_\parallel^2/s \quad (2)$$

and $\Lambda$ is the 2D thin-film screening length; $\lambda_\parallel$ is the effective penetration depth parallel to the graphene planes and *s* is the interlayer spacing.

Where the layers *i* and *j* are adjacent and the vortices are directly vertically aligned, this reduces to:

$$\mathbf{F}(\mathbf{z},0,1) \approx \hat{z}\frac{\varphi_0^2 s^2}{16\pi^2 \lambda_\parallel^4} \quad (3)$$

It is obvious from this simplistic analysis that the magnetic pinning force between two pancake vortices in adjacent layers is proportional to the square of the interlayer spacing and inversely proportional to the fourth power of the magnetic field penetration depth within the layer. Given that the interlayer spacing in $Bi_2Sr_2CaCu_2O_{8+x}$ is 1.5 nm and for graphene stacks it is 0.2 nm, the pinning force in the graphene stacks would be 56 times less than that of $Bi_2Sr_2CaCu_2O_{8+x}$ even if the magnetic field penetration depths were identical. This is significant as it implies that, in the absence of external pinning, flux-flow resistance may be a factor in the resistance of graphene to very low temperatures.

The single pancake vortex pinning energy is given by Clem [22] as:

$$U_0 = \left(\phi_0/4\pi\right)^2 s/\lambda_{ab}^2 \tag{4}$$

where $\lambda_{ab}$ is the in-plane penetration depth.

According to Clem this gives a self-pinning characteristic temperature for a single pancake vortex in $YBa_2Cu_3O_{7-x}$ of 1200 K and vortex motion begins to be a problem at about 1/20 of that temperature. Expressions (5) and (6) below are given for $YBa_2Cu_3O_{7-x}$ [22]:

$$U_0/k_B = 1200 \text{ K} \tag{5}$$

where the flux motion temperature regime is:

$$\left(U_0/k_B\right)/20 = 60 \text{ K} \tag{6}$$

The inter-plane separation of graphene layers is 0.2 nm and that of $YBa_2Cu_3O_{7-x}$ is 1.2 nm so if the in-plane penetration depth $\lambda_{ab}$ is the same the expected temperature where flux motion regime begins to be important for graphene would be about 10 K. Given that the measured conduction anisotropy in graphite is significantly greater than that of $YBa_2Cu_3O_{7-x}$ this would be expected to be an upper value estimate with realistic estimates for the onset temperature for flux flow being much lower. This sets up the expectation that, even if the material is superconducting with Cooper pairs, that the resistance would be non-zero even at low temperatures and currents.

The strength of attraction of a pancake vortex to a vacancy due to implantation damage is related to the size of the vacancy and the in-plane penetration depth that governs the physical size of the pancake vortex. At low temperatures a pancake vortex could be pinned to a vacancy but this pinning force would not be strong in this highly anisotropic material unless there were other pancake vortices pinned to similar vacancies in adjacent layers. If this would be the case then the damage would provide condensation sites for the formation of pancake vortex "stacks". The pinning force of these stacks would be from the summation of the vortices in the stack's mutual magnetic and Josephson interactions as detailed in Clem's work [20]. The temperature required to "melt" one of these stacks is proportional to the pinning energy, primarily from the inter-vortex pinning of the stack, not the vacancy. Stacks with different numbers of pancake vortices in them would thus melt at different temperatures. Once a stack melts all of the vortices in the stack are free to move and immediately contribute to resistive losses in the material. This would lead to the expectation of upwards steps in resistance at the melting temperatures of the various height pancake vortex stacks. Step smearing would be predicted by thermal effects and angular misalignment of some of the vortices in the pancake vortex stacks (tilted stacks of pancake vortices).

It would be remiss of us not to mention the work of others in attempting to understand the resistance versus temperature characteristics of graphite and graphene [28-37]. Generally these efforts have involved the use of several sorts of phonon dispersion, reflection or resonance at a variety of energies in an effort to explain the observed response. The difficulty with such models is that they fail to adequately predict a nearly periodic series of steps and that for each feature observed a new mechanism must be proposed. Finally, none of the phonon-based models can accommodate nearly vertical steps in resistance.

## 3. Experimental

The experimental setup consists of a closed cycle helium refrigerator with a Keithley 6221/2182A Delta Measurement System running under computer control using our custom written LabView data acquisition program to control the entire run. Temperature was likewise computer monitored using a Lakeshore 335 thermometer. All measurements were averaged delta mode in nature and each data point recorded is the result of 250 individual measurements averaged together. Measurement speed is approximately 10 seconds per final averaged data point recorded. All measurements were made at the minimum practical stimulus current, always 10 µA or less, to avoid current induced degradation of the responses in these very thin films. The maximum thermal slew rate for the sample warming up was lower than 0.008 K/s for temperatures greater than 10 K.

All resistance measurements were taken in a four-point configuration with spring-loaded contacts as shown in figure 1.

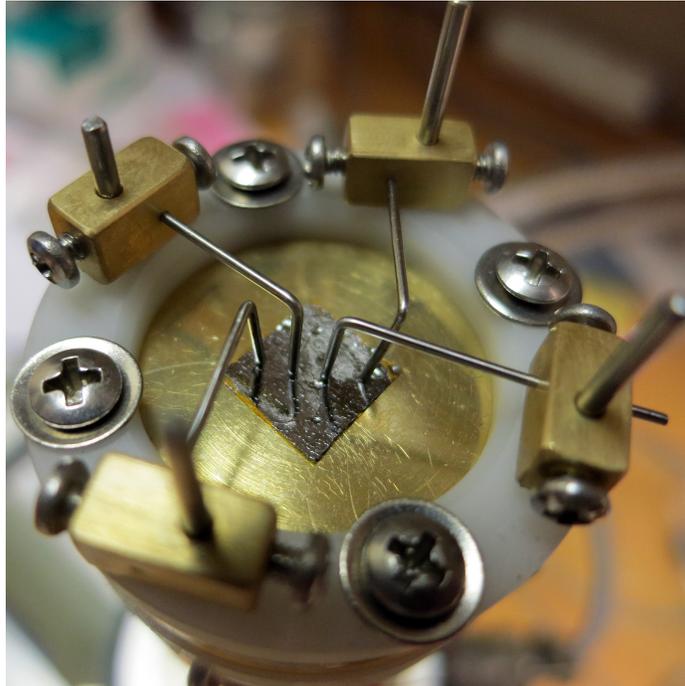

**Figure 1.** Thin film graphite specimen placed in the four-probe test fixture for $R$ vs. $T$ measurement.

The as-purchased grade ZYH MikroMasch or NT-MDT HOPG was first measured in a closed cycle refrigeration system and our well-tested four-point probe resistance vs. temperature ($R$ vs. $T$) system was connected to it.

After this initial testing the samples were then implanted with phosphorus or boron. The implantation of the phosphorous and argon was performed by Cutting Edge Ions, LLC on a mail-in basis. Energy of implanted phosphorus was 10 keV with doses ranging from $6\times10^7$ cm$^{-2}$ to $4\times10^9$ cm$^{-2}$. Argon was implanted with 5 keV energy and dose $1.2\times10^8$ cm$^{-2}$. Overall, more than fifty HOPG samples have undergone ion implantation treatment. The computed depth profile of ion implanted phosphorous in graphite and the computed damage profile are shown in figure 2, curves 1 and 2. The corresponding computed profiles for the ion implanted argon in graphite are shown in figure 2, curves 3 and 4. Since there is no characterized implantation model for the stopping power of HOPG as a substrate we selected the similar substrate material, graphite, for the simulation. For the task of estimating the range, damage and ion distribution in the HOPG the results should be accurate to within a few percent.

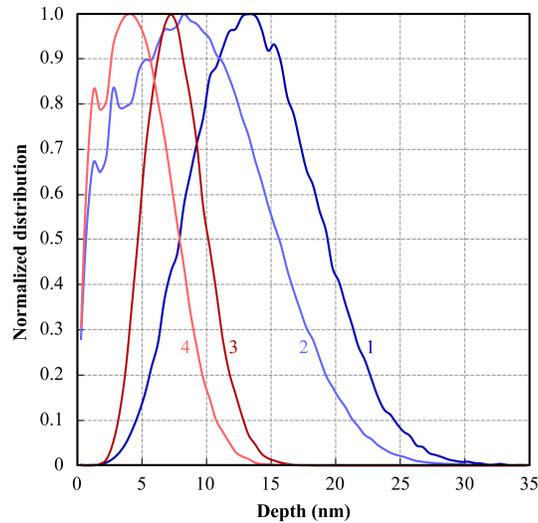

**Figure 2.** Simulated normalized distribution of implanted atoms and lattice damage caused by implants in depth of HOPG. (1) – distribution of implanted Phosphorous; (2) – damage caused by implanted Phosphorous; (3) – distribution of implanted Argon; (2) –damage caused by implanted Argon.

After implantation the sample was again placed in the helium refrigeration system and the *R* vs. *T* characteristic of the HOPG was measured. Once again all measurements were averaged delta mode in nature and each data point recorded is the result of 250 individual measurements averaged together. Measurement speed is approximately 10 seconds per final averaged data point recorded. All measurements were made at the minimum practical stimulus current, always 10 µA or less (typically 1 µA), to avoid current induced degradation of the responses in these very thin films.

After the measuring of the *R* vs. *T* characteristic of the post implantation bulk sample the implanted surface was exfoliated using silicon adhesive Kapton® film tape to get multilayers of graphene. These multilayer graphene films, still adhered to the tape, were then placed into the closed cycle refrigeration system and the *R* vs. *T* characteristic of the graphene multilayer stack was measured. To check that the graphene exfoliations were affected by an applied magnetic field similarly to the bulk phosphorous ion implanted HOPG shown in figure 3 a modest, calculated, 100 mT to 350 mT magnetic field was supplied by a dc current driven coil placed externally over the refrigerator vacuum shroud and the *R* vs. *T* characteristic was re-measured on a number of samples.

## 4. Results
Our prior work [5] showed that phosphorous implanted HOPG samples exhibit a deviation from the expected monotonic rise in resistance as temperature goes down at some point above 100 K (figure 3). This and the fact that there is a fairly steep drop in the resistance (by a factor of more than 2) at lower temperatures was enough to be considered an indicator of possible superconductivity in the sample.

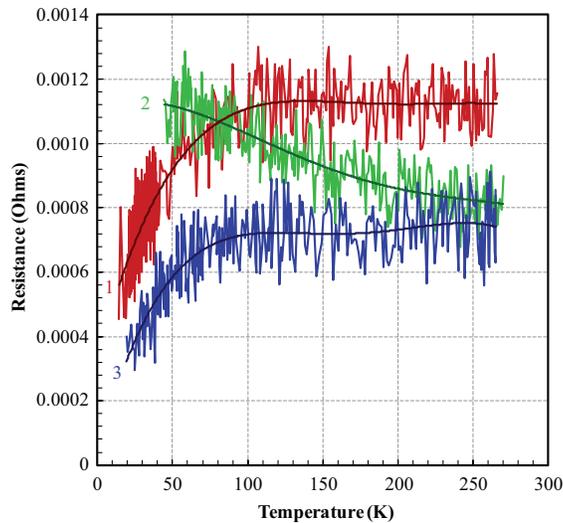

**Figure 3.** Measured *R* vs.*T* dependence of Phosphorus-doped bulk HOPG-005 sample. (1) before magnetic field was applied. (2) with magnetic field applied. (3) after magnetic field was removed. Smooth lines are a fourth order polynomial fit.

Dr. Paul Bach of Dr. Richard Greene's group at the University of Maryland measured the magnetic field characteristics of the same phosphorous implanted bulk HOPG sample; this is shown in figure 4.

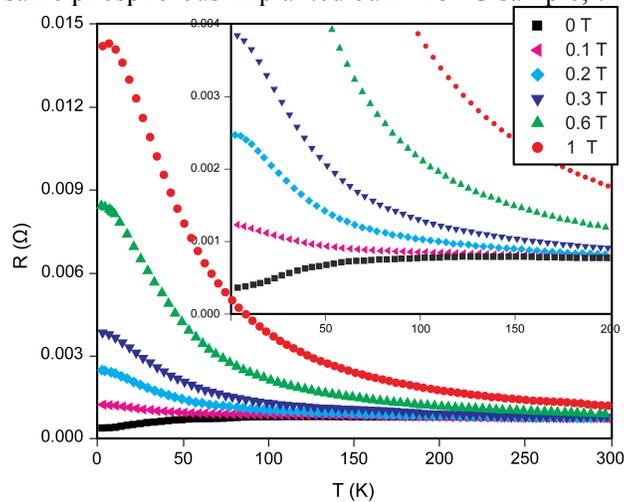

**Figure 4.** *R* vs. *T* measurement of Phosphorus doped HOPG sample 005 in the presence of magnetic field from 0 T to 1 T.

The lack of zero resistance and the modest magnetic field (maximum attainable was less than 0.035 T) required to quench the effect even in the exfoliated multilayer graphene samples is shown in figure 5 for a representative sample.

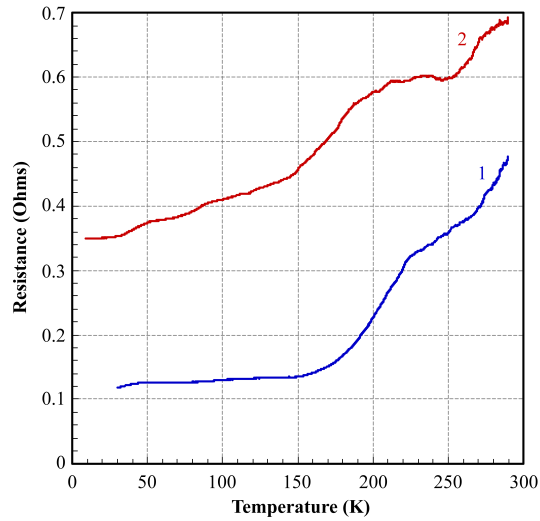

**Figure 5.** *R* vs. *T* of thin film exfoliated from Phosphorus doped HOPG sample 008 measured without (1) and with 0.035 T magnetic field applied (2).

The natural inclination is to look first at the high temperature features in this *R* vs. *T* characteristic, however, the key to understanding the potential causes for the observed results are best understood by examining a number of *R* vs. *T* characteristics, shown in figure 6, of similarly exfoliated films taken from bulk HOPG implanted using phosphorous ions.

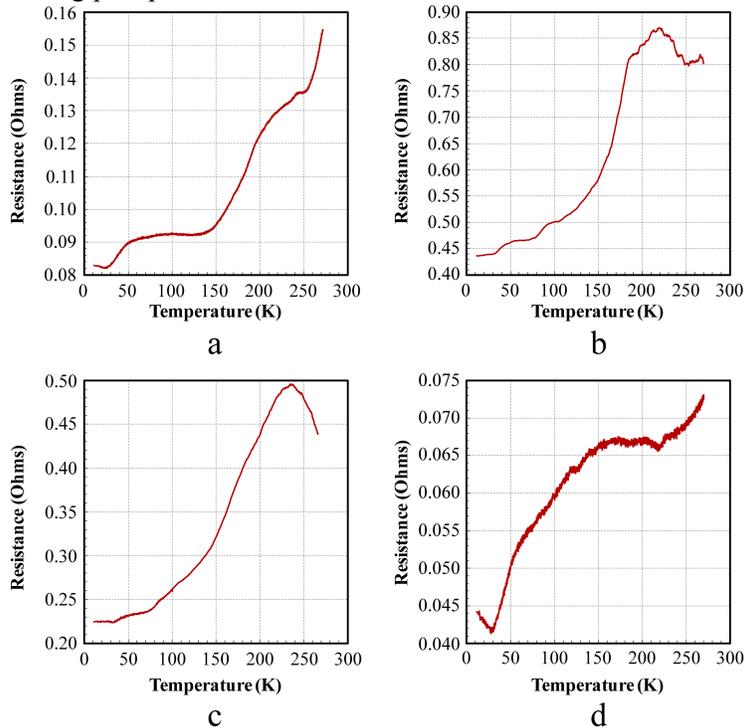

**Figure 6.** Measured *R* vs *T* of four thin films exfoliated from phosphorus implanted ($E_p$ = 10 keV, dose $1.2 \times 10^8$ cm$^{-2}$) HOPG samples. (a) – HOPG-008, layer 3; (b) – HOPG-019, layer 3; (c) – HOPG-019, layer 6; (d) – HOPG-021, layer 7.

Comparing characteristics a-d in figure 6 it is clear that there is a step in resistance, in all of the samples, that occurs at a temperature of approximately 50-60 K. What is likewise clear when the data is closely examined is that there is a second resistance step at 100-120 K. Again, at a temperature of from approximately 150-180 K and, yet again at a temperature from about 200-240 K there are additional steps in the $R$ vs. $T$ characteristics of all or almost all of the samples. This alone is cause to suspect that the features are due to magnetic vortex lattice melting and subsequent flux flow losses.

In an effort to see if additional lattice damage by neutral ion species could reduce the losses one sample (HOPG-023), which had been previously implanted with phosphorous but had not yet been exfoliated, was sent back for implantation with argon. This implantation was done at reduced energy and the same dose (5 keV and $1.2\times10^8$ cm$^{-2}$) to place the damage in front of the peak in the phosphorous distribution as shown in figure 2 for computed range distributions of the phosphorous (1) and the argon (3) implants in depth of HOPG sample and for computed damage distributions caused by implanted phosphorous (2) and argon (4). A change in the $R$ vs. $T$ characteristic that was qualitatively commensurate with stronger pinning would indicate that the resistance was due to flux flow.

The $R$ vs. $T$ characteristic of an exfoliated graphene multilayer from this doubly implanted sample, HOPG-023, is shown in the curve 1 in figure 7. Note that the first $R$ vs. $T$ characteristic taken showed the same qualitative behaviour that the samples in figure 6.

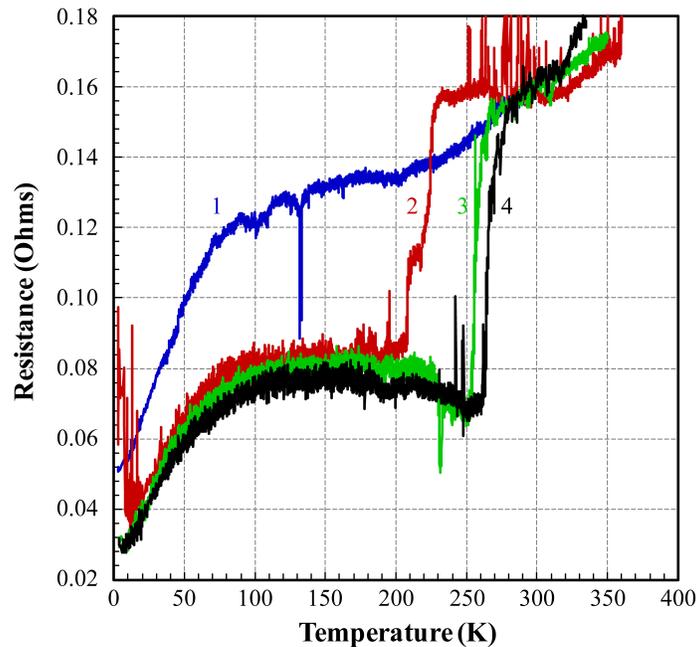

**Figure 7.** $R$ vs. $T$ of thin film sample peeled off Phosphorous-implanted and then Argon-implanted bulk HOPG-023. Curves 1 to 4 are four identical sequential runs with the same probe position.

A small notch in curve 1 (figure 7), somewhat too wide (18 data points each from 250 averaged measurements) to be either noise or a measurement glitch, at a temperature of 132 K caused the sample to be re-measured multiple times. These re-measurements were performed without moving the contacts or opening the refrigeration system. In these subsequent $R$ vs. $T$ re-measurements, shown as curves 2 through 4 in figure 7, a large and abrupt step was observed. The step was in the temperature region of 210-230 K on the first re-measurement. On a second re-measurement it was observed to have moved upward to a temperature of 250-260 K. On a third, and final, re-measurement, the step was noted to have migrated upwards to a temperature of 264-267 K.

When all four $R$ vs. $T$ characteristics for the graphene multilayer from sample HOPG-023 are plotted together on the same graph (figure 7) it is clear that these steps are not a contact issue in the measuring system as the low temperature and the high temperature resistances have not been changed significantly from run to run. It is also clear that the notch observed in the first $R$ vs. $T$ characteristic is an attempt for the material to move to the resistance state achieved in the later runs and is not to be dismissed as spurious.

Finally, it was noted that the use of lower currents gave more distinct and larger steps; larger currents often showed only one or two steps whereas a lower current would reveal several additional steps. This is also consistent with possible superconductivity as the current densities involved with a microampere in our material would be in excess of 10 A/cm$^2$ (10$^5$ A/m$^2$) for a 1 nm thick film if the entire cross section were carrying current. Given a dose of ~2×10$^8$ cm$^{-2}$, a surface atomic density of 10$^{14}$ cm$^{-2}$ and a thickness of ~5 atomic layers this puts the volume ratio of implanted donors to carbon atoms at 4×10$^{-7}$ to 1. As a result local current densities may be in the 10$^6$ A/cm$^2$ to 10$^7$ A/cm$^2$ range depending on the effectiveness of the donors in activating states.

## 5. Discussion

We have observed a response consistent with the presence of magnetic field flux vortices in phosphorous (electron donor) implanted Highly Oriented Pyrolytic Graphite and in phosphorous doped exfoliated multilayer graphene. The repeated nature of the observed steps in the $R$ vs. $T$ characteristics of the material is consistent with the melting of stacks of pancake vortices of differing lengths at different temperatures. The lack of zero resistance at low temperatures is also consistent with pancake vortex behaviour in the flux-flow regime. This allows the use of a single phenomenon, magnetic pancake vortices, to describe the features observed. The presence of magnetic vortices requires, and is direct evidence of, superconductivity. The lack of Meissner effect observation or the failure to detect a superconducting energy gap may indicate that the effect is due to something other than superconductivity. It may also simply mean that the volume fraction of material involved (far under the 100 ppm range that the STM was capable of resolving) was too low to measure.

The material that was subjected to post doping argon implantation (damage) showed a discontinuous step in resistance at a temperature of about 265 K. We noted in our prior work [5], the bulk of which is included in the background portion of this paper, that "the ultimate critical temperature in this system is in excess of 100 K and, may very likely be considerably higher if damage incurred during the doping can be further minimized". Clearly this conclusion has not been voided and may well be valid for a transition at a temperature of greater than 300 K.

## 6. Conclusions

There are two direct conclusions to be drawn from this work: (1) the damage done in the implantation(s) has allowed the effect to be more clearly observed and, (2) the presence of a step in the resistance versus temperature characteristic at 265 K suggests that the actual transition temperature may lie even higher in temperature.

This evidence of superconductivity (or other new physical phenomenon) in doped graphene/graphite with a de-pairing (critical) temperature in the region of 265 K and possibly higher is intriguing. Clearly more work needs to be done to confirm the observations and proposed mechanism of the observed effect. We welcome further investigation into the matter by our colleagues.

**Acknowledgments**
The authors wish to thank Dr. Richard Greene and Dr. Paul Bach at the University of Maryland who tested our samples to confirm the observed Resistance vs. Temperature curves which are shown in this work. This work was supported by the United States Air Force under AFOSR grant FA9550-10-1-0134.